\documentclass[a4paper]{easychair}
\usepackage{geometry}
\usepackage{listings,multicol}
\usepackage{lstcoq}
\usepackage{amssymb}
\usepackage{subcaption}
\usepackage{semantic}
\usepackage[toc]{appendix}
\usepackage[inline]{enumitem}
\usepackage{hyperref}
\usepackage[capitalise,nameinlink,noabbrev]{cleveref}

\usepackage{xcolor}

\definecolor{ltblue}{rgb}{0,0.4,0.4}
\definecolor{dkblue}{rgb}{0,0.1,0.6}
\definecolor{dkgreen}{rgb}{0,0.5,0}
\definecolor{brightmaroon}{rgb}{0.76, 0.13, 0.28}
\definecolor{burntorange}{rgb}{0.8, 0.33, 0.0}
\definecolor{dkred}{rgb}{0.5,0,0}
\definecolor{bggray}{gray}{0.95}
\definecolor{arsenic}{rgb}{0.23, 0.27, 0.29}

\newcommand{\icode}[1]{\lstinline[mathescape,basicstyle=\small]!#1!}

\newtheorem{theorem}{Theorem}

\newtheorem{definition}{Definition}

\newcommand\CIC[0]{CIC}
\newcommand\CICbox[0]{$\lambda\square$}

\reservestyle{\command}{\mathtt}
\command{if[if],then[then],else[else],let[let],in[in],case[case],return[return],of[of],fix[fix],
fun[fun],Set[Set],forall[forall],match[match],as[as],in[in],with[with],end[end],
NotPrenex[NotPrenex],Ok[Ok],TypeError[TypeError],TypingError[TypingError],
Error[Error],NotSupported[NotSupported],Checked[Checked],true[true],false[false],
ret[ret],isarity[is\_arity],islogical[is\_logical],issort[is\_sort],
tSort[Type],RelOther[Other],RelTypeVar[TV],RelInductive[Ind],
tConst[C],tInd[I]}

\colorlet{cicterms}{dkblue}
\colorlet{boxtyterms}{dkgreen}
\colorlet{errorcolor}{red}

\newcommand\cic[1]{{\color{cicterms}{#1}}}

\newcommand\boxty[1]{{\color{boxtyterms}{#1}}}
\newcommand\TBox{\boxty{\square}}
\newcommand\TAny{\boxty{\ensuremath{\mathbb{T}}}}
\newcommand\flagoftype{{\ident{flag\_of\_type}}}

\newcommand\ident[1]{{\mathit{#1}}}
\newcommand\erasetypename{\ensuremath{\mathcal{E}^T}}
\newcommand\erasetype[4]{\erasetypename{}~#1~#2~#3~#4~}
\newcommand\erasetypeappname{\mathcal{E}_{app}^T}
\newcommand\erasetypeapp[4]{\erasetypeappname{}~#1~#2~#3~#4~}
\newcommand\erasetypeheadname{{\ensuremath{\mathcal{E}_{head}^T}}}
\newcommand\erasetypehead[2]{\erasetypeheadname{}~#1~#2}
\newcommand\erasevarname{{\ensuremath{\mathcal{E}_{var}^T}}}
\newcommand\erasevar[2]{{\erasevarname{}~#1~#2}}
\newcommand\type[1]{\ensuremath\mathtt{#1}}

\newcommand\dearg{{\ident{dearg}}}
\newcommand\deargcst{{\ident{dearg\_cst}}}
\newcommand\deargmib{{\ident{dearg\_mib}}}
\newcommand\deargenv{{\ident{dearg\_env}}}

\newcommand\wcbveval[3]{{\ensuremath{#1 \vdash #2 \triangleright #3}}}
\newcommand\wcbvevalpcuic[3]{{\ensuremath{#1 \vdash_p #2 \triangleright #3}}}

\newcommand\error[1]{{\color{errorcolor}{#1}}}
\newcommand\redbetaiotazeta{{\mathtt{red}_{\beta\iota\zeta}}}
\newcommand\decomposeapp{{\mathtt{decompose\_app}}}

\newcommand{\ARXIV}{} 

\title{Extracting Smart Contracts Tested and Verified in Coq}
\titlerunning{Extracting Smart Contracts Tested and Verified in Coq}

\author{Danil Annenkov\inst{1} \and Mikkel Milo\inst{2} \and Jakob Botsch Nielsen\inst{1} \and Bas Spitters\inst{1}}
\institute{Concordium Blockchain Research Center, Aarhus University \and Department of Computer Science, Aarhus University, Denmark}
\authorrunning{Annenkov, Milo, Nielsen, Spitters}

\begin{document}

\maketitle

\begin{abstract}
  We implement extraction of Coq programs to functional languages based on MetaCoq's certified erasure.
  As part of this, we implement an optimisation pass removing unused arguments.
  We prove the pass correct wrt.\ a conventional call-by-value operational semantics of functional languages.
  We apply this to two functional smart contract languages, Liquidity and Midlang, and to the functional language Elm.
  Our development is done in the context of the ConCert framework that enables smart contract verification.
  We contribute a verified boardroom voting smart contract featuring maximum voter privacy such that each vote is kept private except under collusion of all other parties.
  We also integrate property-based testing into ConCert using QuickChick and our development is the first to support testing properties of interacting smart contracts.
  We test several complex contracts such as a DAO-like contract, an escrow contract, an implementation of a Decentralized Finance (DeFi) contract which includes a custom token standard (Tezos FA2), and more.
  In total, this gives us a way to write dependent programs in Coq, test them semi-automatically, verify, and then extract to functional smart contract languages, while retaining a small trusted computing base of only MetaCoq and the pretty-printers into these languages.
\end{abstract}

\section{Introduction}\label{sec:intro}
Smart contracts are programs running on top of a blockchain.
They often control large amounts of cryptocurrency and cannot be changed after deployment.
Unfortunately, many vulnerabilities have been discovered in smart contracts and this has led to huge financial losses (e.g.\ TheDAO, Parity's multi-signature wallet).
Therefore, smart contract verification is crucially important.
Functional smart contract languages are becoming increasingly popular: e.g.\ Simplicity~\cite{O'Connor:Simplicity}, Liquidity~\cite{Liquidity}, Plutus~\cite{Chapman:PlutusCore}, Scilla~\cite{Sergey:ScillaOOPSLA} and Midlang\footnote{\url{https://developers.concordium.com/midlang}}.
A contract in such a language is just a function from a message type and a current state to a new state and a list of actions (transfers, calls to other contracts), making smart contracts more amenable for formal verification.
Functional smart contract languages, similarly to conventional functional languages, are often based on a variants of System F allowing the type checker to catch many errors. For errors that are not caught by the type checker, a proof assistant, in particular Coq, can be used to ensure correctness.

Formal verification is a complex and time-consuming activity, and much time may be wasted attempting to prove false statements (e.g.\ if the implementation is incorrect).
Property-based testing is an automated testing technique with high bug-discovering capability compared to e.g.\ unit testing.
It can provide a preliminary, cost-efficient approach to discover implementation bugs in the contract or mistakes in the statement to be proven.

Once properties of contracts are tested and proved correct, one would like to execute them on blockchains.
One way of achieving this is to \emph{extract} the executable code from the formalised development.
Various verified developments rely on the extraction feature of proof assistants extensively~\cite{2006-Leroy-compcert,klein14tocs,ExtractionLargeProofs:CrusFilipeSpitters,ExecutingExtracted:CruzFilipeLetouzey,FSets:FillaitreLetozey}.
However, currently, the standard extraction feature in proof assistants supports conventional functional languages (Haskell, OCaml, Standard ML, Scheme, etc.) by using unsafe features such as type casts, if required, and this is not possible in many smart contracts languages.
More importantly, the current implementation of extraction is written in OCaml and is not verified.
\begin{figure}
  \centering
  \includegraphics[width=8.5cm]{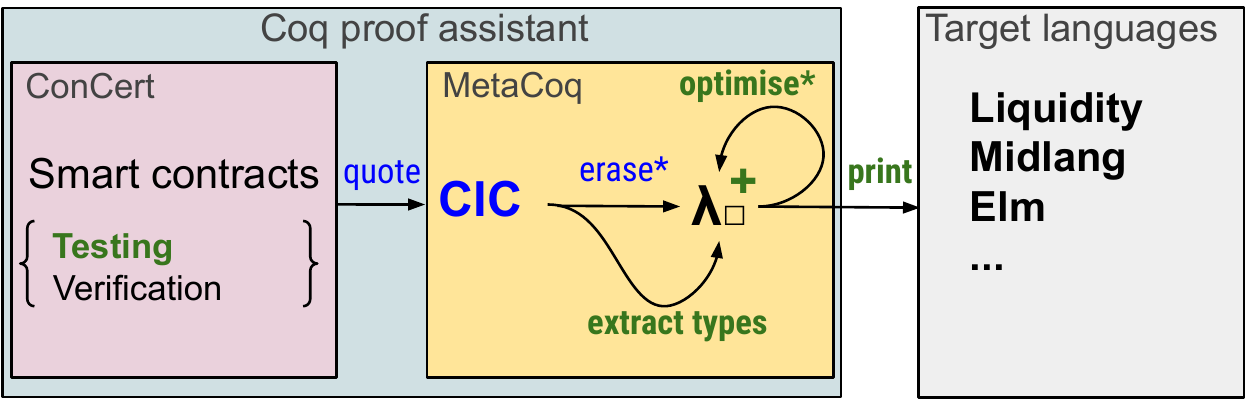}
  \caption{The pipeline}\label{fig:pipeline}
\end{figure}

\paragraph{Contributions}
We build on the ConCert framework~\cite{ConCert} for embedding smart contracts in Coq and the execution model introduced in~\cite{Interactions}.
We summarise the contributions of this work as the following.
\begin{itemize}
\item We provide a general framework for extraction from Coq to a typed functional language (\cref{sec:extraction}).
  The framework is based on certified erasure~\cite{CertErasure}, which we extend with an erasure procedure for types and inductive definitions.
  Moreover, we implement and prove correct an optimisation procedure that removes unused arguments and allows therefore for optimising away some computationally irrelevant bits left after erasure.
  We develop the pretty-printers for printing code into two smart contract languages: Liquidity (Tezos and Dune networks) and Midlang (Concordium network).
  Since Midlang is a fork of the Elm functional language~\cite{ElmInAction}, our extraction also supports Elm as a target language.
\item We integrate contract verification and testing using QuickChick, a property-based testing framework for Coq, by testing properties on generated execution traces.
  We show that this would have allowed us to detect several high profile exploits automatically.
  The testing frameworks itself uses the facilities of Coq for giving stronger correctness guarantees for the testing infrastructure (\cref{sec:testing}).
\item We provide case studies of smart contracts in ConCert by testing and proving properties of an escrow contract and an anonymous voting contract based on the Open Vote Network protocol (\crefrange{sec:escrow}{sec:boardroom}).
    We apply our extraction functionality to study the applicability to the developed contracts.
\end{itemize}
Our work is the first development applying the property-based testing techniques to smart contract \emph{execution traces} (as opposed testing single-step executions), and extracting the verified programs to smart contract languages.

The pipeline covering the full process of starting with a smart contract as a function in Coq and ending with extracted code in one of the target languages is shown in \cref{fig:pipeline}.
The \textbf{\boxty{green}} items are contributions of this work and the items marked with $^{*}$ are verified in Coq.
The MetaCoq project~\cite{MetaCoq} provides us with metaprogramming facilities (e.g.\ quoting Coq terms) and formalisation of the meta-theory of Coq (a variant of the calculus of inductive constructions --- CIC).
By \CICbox$^\boxty{+}$, we mean the untyped calculus of extracted programs (part of MetaCoq) enriched with data structures required for type extraction.

Our trusted computing base (TCB) includes Coq itself, the quote functionality of MetaCoq and the pretty-printing to target languages.
While the type extraction procedure is not verified, it does not affect the soundness of the pipeline (see discussion in \cref{sec:smart-contract-extraction}).

Our development is available as an accompanying artifact~\cite{ConCertArtifact} and in the GitHub repository~\url{https://github.com/AU-COBRA/ConCert/tree/artifact-2020}.

\section{The ConCert Framework}\label{sec:ConCert}
The present work builds on and extends the ConCert smart contract certification framework~\cite{ConCert}.
The ConCert framework features an embedding of smart contracts into Coq along with the proof of soundness of the embedding using the MetaCoq project~\cite{MetaCoq}.
The embedded contracts are available in the deep embedding (as ASTs) and in the shallow embedding (as Coq functions).
Having smart contracts as Coq functions facilitates the reasoning about their functional correctness properties.
Moreover, ConCert features an execution model first introduced in~\cite{Interactions}.
The execution model allows for reasoning on contract execution traces which makes it possible to state and prove temporal properties of interacting smart contracts.
The previous work on ConCert~\cite{ConCert} mainly concerns with the following use-case: take a smart contract, \emph{embed} it into Coq and verify its properties.
This work explores how it is possible to verify a contract as a Coq function and then \emph{extract} it into a program in a functional smart contract language.

\section{Extraction}\label{sec:extraction}
The Coq proof assistant features a possibility of extracting the executable content of Gallina terms into OCaml, Haskell and Scheme~\cite{letouzey04}.
This functionality thus enables proving properties of functional programs in Coq and then automatically producing code in one of the supported languages.
The extracted code can be integrated with existing developments or used as a stand-alone program.
The extraction procedure is non-trivial since Gallina is a dependently typed functional language.
Recent projects such as MetaCoq~\cite{CertErasure} and CertiCoq~\cite{Anand:CertiCoq} provide formal guarantees that the extraction procedure is correct, but do not support extraction to smart contract languages.
The general idea of extraction is to find and mark all parts of a program that do not contribute to computation.
That is, types and propositions in terms are replaced with $\square$~(a box).
Formally, it is expressed as a translation from \CIC{} (Calculus of Inductive Construction --- the underlying calculus of Coq) to \CICbox{}~\cite{letouzey04,CertErasure}.
In the present work, by \CIC{} we mean the predicative cumulative calculus of inductive constructions (pCuIC), as presented by the authors of~\cite{CertErasure}.
\CICbox{} is an untyped version of \CIC{} with an additional constant $\square$.
One of the important results of~\cite{letouzey04} is that the computational properties of the erased terms are preserved.
That assumes that the extracted code is untyped, while integration with the existing functional languages requires to recover the typing.
In~\cite{letouzey04} this problem is solved by designing an extraction procedure for types and then using the modified type inference algorithm (based on the algorithm $\mathcal{M}$~\cite{AlgM:LeeYi}) to recover types and check them against the type produced by extraction.
Because the type system of Coq is more powerful than type systems of the target languages (e.g.\ Haskell or OCaml) not all the terms produced by extraction will be typable.
In this case the modified type inference algorithm inserts type coercions forcing the term to be well-typed. If we step a bit outside the OCaml type system (even without using dependent types), the extraction will have to resort to \icode{Obj.magic} in order to make the definition well-typed.
For example, the code snippet below
\begin{lstlisting}
Definition rank2 : forall (A : Type), A->(forall A : Type, A->A)->A
  := fun A a f => f _ a.
Extraction rank2.
\end{lstlisting}
gives the following output on extraction to OCaml:
\begin{lstlisting}
(** val rank2 : 'a1 -> (__ -> __ -> __) -> 'a1 **)
let rank2 a f = Obj.magic f __ a
\end{lstlisting}
\noindent%
These coercions are ``safe'' in the sense that they do not change the computational properties of the term, they merely allow to pass the type checking.

Since the extraction implementation becomes part of a TCB, one would like mechanically verify the extraction procedure in Coq itself.
An important step in this direction was made by the MetaCoq project~\cite{MetaCoq}, which includes \emph{certified erasure}~\cite{CertErasure} that specifies an erasure procedure as a translation to \CICbox{} in Coq and proves that the evaluations of original and erased terms agree.

\lstset{keepspaces = true}
\subsection{Extraction to Functional Smart Contract Languages}\label{sec:smart-contract-extraction}
We target functional smart contract languages, which often pose more challenges than the conventional targets for extraction.\footnote{Our implementation of the extraction procedure is available in the \texttt{extraction} subfolder of the artifact.}
We have identified the following restrictions.
\begin{enumerate}
\item Most of the smart contract languages\footnote{At least, Simplicity, Liquidity, CameLIGO (and other LIGO languages), Love, Scilla, Sophia, Midlang.} do not offer a possibility to insert type coercions forcing the type checking to succeed.
\item The operational semantics of \CICbox{} has the following rule (see Section 4.1 in~\cite{CertErasure}): if $\wcbveval{\Sigma}{t_1}{\square}$ and $\wcbveval{\Sigma}{t_2}{\ident{v}}$ then $\wcbveval{\Sigma}{\bigl(t_1~~t_2 \bigr)}{\square}$, where $\wcbveval{-}{-}{-}$ is a big-step evaluation relation for \CICbox{}, $t_1$ and $t_2$ are \CICbox{} terms, and $\ident{v}$ is a \CICbox{} value.
  This rule can be implemented in OCaml using the unsafe features, which are, again, not available in most of the smart contract languages.
  In lazy languages this situation never occurs (see Section 2.6.3 in~\cite{letouzey04}), but most languages for smart contracts follow the eager evaluation strategy.
\item Data types and recursive functions are often restricted. E.g.\ Liquidity, CameLIGO (and other LIGO languages) do not allow for defining recursive data types (like lists and trees) and limits recursive definitions to tail recursion on a single argument.\footnote{Some languages do not have this restriction, e.g.\ Midlang and Love.}
  Instead, these languages offer built-in lists and finite maps.
  Scilla exposes only recursors for lists instead of allowing to write recursive functions explicitly.
\end{enumerate}
Regardless of our design choices, the soundness of the extraction (given that terms evaluate in the same way before and after extraction) will not be affected.
In the worst case, the extracted term will be rejected by the type checker of a target language.

At the moment, we consider the formalisation of typing in target languages out of scope for this project.
Even though the extraction of types is not verified, it does not compromise run-time safety as we stated above: if extracted types are incorrect, the target language's type checker will reject the extracted program.
If we followed the work in~\cite{letouzey04}, which the current Coq extraction is based on, giving guarantees about typing would require formalising of target languages type systems, including a type inference algorithm (possibly algorithm $\mathcal{M}$~\cite{AlgM:LeeYi}).
The type systems of many languages we consider are not precisely specified and are largely in flux.
Moreover, for the target languages without unsafe coercions, some of the programs will be untypeable in any case.
Therefore, we do our best to extract as many programs that pass type checking as possible or fail at the extraction stage, due to incompatibilities with the ``generic'' type system usually found in functional languages (we take prenex-polymorphic System F for that purpose).

On the other hand, for more mature languages (e.g.\ Elm) one can imagine connecting our formalisation of extraction with the language formalisation, proving the correctness statement for both the run-time behaviour and the typeability of extracted terms.

Let us consider in detail what the restrictions outlined above mean for extraction.
The first restriction means that certain types will not be extractable.
Therefore, our goal is to identify a practical subset of extractable Coq types.
The second restriction is hard to overcome, but fortunately, this situation should not often occur on the fragment we want to work.
Moreover, as we noticed before, terms that might give an application of a box to some other term will be ill-typed and thus, rejected by the type checker of a target language.
The third restriction can be addressed by mapping Coq's data types (lists, finite maps) to the corresponding primitives in a target language.

We extend the work on certified erasure~\cite{CertErasure} and develop an approach that uses a minimal amount of unverified code that can affect the soundness of the certified erasure.
Our approach adds an erasure procedure for types, simple verified optimisations of the extracted code and pretty-printers for target smart contract languages.

Before introducing our approach, let us give some examples of how the certified erasure works and motivate the optimisations we propose.

\begin{lstlisting}
Definition sum_nat (xs : list nat) : nat := fold_left plus xs 0.
\end{lstlisting}
produces the following \CICbox{} code:
\begin{lstlisting}
fun xs => Coq.Lists.List.fold_left  ∎ ∎  Coq.Init.Nat.add xs O
\end{lstlisting}
Where the \lstinline{∎} symbol corresponds to computationally irrelevant parts.
The first two arguments to the erased versions of \icode{fold_left} are boxes, since \icode{fold_left} in Coq has two implicit arguments. They become visible if we switch on printing of implicit arguments:
\begin{lstlisting}
Set Printing Implicit.
Print sum_nat.
(* fun xs:list nat=>@fold_left nat nat Init.Nat.add xs 0 *)
\end{lstlisting}
In this situation we have at least two choices: remove the boxes by some optimisation procedure, or leave the boxes and extract \icode{fold_left} in such a way that the first two arguments belong to some dummy data type.\footnote{There are two more rules in the semantics of \CICbox{} that do not quite fit into the evaluation model of smart contract languages: pattern-matching on a box argument and having a box as an argument to a fixpoint.
The matching on $\square$ occurs when eliminating from logical inductive types with no constructors (e.g.\ \icode{False}) or from singleton types (e.g. equality type). A special rule for fixpoints is needed because of logical argument to fixpoints used by the accessibility predicate. We address the \icode{False} case in an ad hoc way at the end of \cref{sec:midlang-extraction}. We believe that it is possible to address other cases similarly to the previous works on extraction (Section 4~\cite{CoqNewExtraction} and Section 2.6 in~\cite{letouzey04}), apart from the implementation of $\square$ as an argument consuming function, due to the absence of unsafe features.}
The latter choice cannot be made for some smart contract languages due to restrictions, therefore, we have to remap \icode{fold_left} and other functions on lists to the corresponding primitive functions.
In the following example,
\begin{lstlisting}
Definition square (xs : list nat) : list nat := map (fun x => x * x) xs.
\end{lstlisting}
\noindent
the \icode{square} function erases to
\begin{lstlisting}
fun xs => Coq.Lists.List.map ∎ ∎ (fun x => Coq.Init.Nat.mul x x) xs
\end{lstlisting}
\noindent
The corresponding language primitive would be a function with the following signature: \icode{TargetLang.map: ('a -> 'b) -> 'a list -> 'b list}.
Clearly, there are two extra boxes in the extracted code that prevent us from directly replacing \icode{Coq.Lists.List.map} with \icode{TargetLang.map}.
Instead, we would like to have the following:
\begin{lstlisting}
  fun xs => Coq.Lists.List.map (fun x => Coq.Init.Nat.mul x x) xs
\end{lstlisting}
\noindent
In this case, we can provide a translation table to the pretty-printing procedure mapping \icode{Coq.Lists.List.map} to \icode{TargetLang.map}.
Alternatively, if one does not want to remove boxes, it is possible to implement a more sophisticated remapping procedure. It could replace \icode{Coq.Lists.List.map ∎ ∎} with \icode{TargetLang.map}, but it would require finding all constants applied to the right number of arguments (or $\eta$-expand constants) and only then replace them.
Remapping of inductive types in the same style would involve more complications: constructors of polymorphic inductives will have an extra argument of a dummy type. This would require more complicated pretty-printing of pattern-matching in addition to the similar problem with extra arguments on the application sites.
\lstset{keepspaces = false}

By choosing to implement the optimisation procedure we achieve two goals: remove redundant computations and make the remapping easier.
Removing the redundant computations is beneficial for smart contract languages, since it decreases the cost of a computation in terms of \emph{gas}.
Users typically pay for calling smart contracts and the price is determined by the gas consumption.
That is, gas serves as a measure of computational resources required for executing a contract.
It is important to emphasise that we can pretty-print code produced by the certified erasure procedure directly.
Moreover, it is important to separate these two aspects of extraction: erasure (given by the translation \CIC{} $-->$ \CICbox{}) and optimisation of \CICbox{} terms to remove unnecessary arguments.
The optimisations we propose are simple, make the output more readable and facilitate the remapping to the target language's primitives.

Our implementation strategy of extraction is the following:
\begin{enumerate*}[label=(\roman*)]
\item take a term and erase it and its dependencies recursively to get an environment;
\item analyse the environment to find optimisable types and terms;
\item optimise the environment in a consistent way;
\item pretty-print the result in the target language syntax.
\end{enumerate*}

\paragraph{Erasure for Types}
Let us discuss our first extension to the certified erasure presented in~\cite{CertErasure}, namely an \emph{erasure procedure for types}.
It is a crucial part for extracting to a \emph{typed} target language.
\begin{figure*}
  \small{\begin{equation*}
      \begin{aligned}[t]
        \erasetypename{} & : \type{Ctx} -> \type{ECtx} -> \cic{\type{term}} -> \type{list~name} \\ &
        -> \type{result}~(\type{list~name} \times \boxty{\type{box\_type}})\\
        \ifdefined\ARXIV{}
          \erasetype{&\Gamma}{\Gamma_e}{\ident{\cic{t}}}{\ident{vs}} := \<let>~\ident{\cic{t'}} := \\
          & \redbetaiotazeta~\Gamma~\ident{\cic{t}}~\<in>\\
        \else
          \erasetype{\Gamma}{\Gamma_e}{\ident{\cic{t}}}{\ident{vs}} & := \<let>~\ident{\cic{t'}} := \redbetaiotazeta~\Gamma~\ident{\cic{t}}~\<in>\\
        \fi
        & \ident{flag} <- \flagoftype~\Gamma~\ident{\cic{t'}};\\
        & \<if>~(\<islogical>~\ident{flag})~\<then>~\<Ok>~\TBox ~ \<else>\\
        & \<match>~\ident{t'}~\<with>\\
        & |~\cic{\overline{i}} => Ok (\ident{vs},\erasevar{\Gamma_e}{i})\\
        & |~\cic{\<tSort>} => \<Ok>~\TBox\\
        & |~\cic{\<forall>~ a : A, B} =>\\
        & \quad \ident{flag} \leftarrow \flagoftype~\Gamma~\cic{A};\\
        & \quad \<if>~(\<islogical>~\ident{flag})~\<then>\\
        & \qquad (\ident{vs_\tau},\boxty{\tau}) <- \erasetype{(\ident{\cic{A}}::\Gamma)}{(\<RelOther>::\Gamma_e)}{\cic{B}}{\ident{vs}}; \\
        & \qquad \<Ok>~(\ident{vs_\tau}, \boxty{\square --> \tau})\\
        & \quad \<else>~\<if>~not(\<isarity>~\ident{flag})~\<then>\\
        & \qquad (\ident{vs_\sigma}, \boxty{\ident{\sigma}}) <- \erasetype{\Gamma}{\Gamma_e}{\ident{A}}{\ident{vs}} ;\\
        & \qquad\<if>~(|\ident{vs}| < |\ident{vs_\sigma}|)~\<then>~\error{\<NotPrenex>} \\
        & \qquad \<else>~(\ident{vs_\tau},\boxty{\tau}) <- \erasetype{(\ident{\cic{A}}::\Gamma)}{(\<RelOther>::\Gamma_e)}{\cic{B}}{\ident{vs}}; \\
        & \qquad\quad \<Ok>~(\ident{vs_\tau}, \boxty{\sigma --> \tau})\\
        & \quad \<else>~\<if>~(\<issort>~\ident{flag})~\<then>\\
        & \qquad\quad(\ident{vs_\tau},\boxty{\tau}) <- \erasetype{(\ident{\cic{A}}::\Gamma)}{(\<RelTypeVar>|\ident{vs}| :: \Gamma_e)}{\cic{B}}{(\ident{vs}+\!\!\!+[\cic{a}])}; \\
        & \qquad\quad \<Ok>~(\ident{vs_\tau}, \boxty{\square --> \tau})\\
        & \quad \<else>~\error{\<NotPrenex>}\\
        & \quad |~ \cic{\bigl(\ident{u}~~\ident{v}\bigr)} => \<let> (\cic{\ident{hd}},\cic{\ident{args}}):= \decomposeapp~\cic{\left(\ident{u}~~\ident{v}\right)}~\<in>\\
        & \qquad \boxty{\sigma} <-\erasetypehead{\Gamma_e}{\cic{\ident{hd}}};\\
        & \qquad\erasetypeapp{\Gamma_e}{\cic{\ident{args}}}{\ident{vs}}{\boxty{\sigma}}\\
        & \quad|~ \cic{\<tConst>} => \<Ok>~(\ident{vs},\boxty{\<tConst>})
         \quad|~ \cic{\<tInd>} => \<Ok>~(\ident{vs},\boxty{\<tInd>})\\
        & \<end>
      \end{aligned}
      \begin{aligned}[t]
        \erasetypeappname{} : \type{ECtx}& -> \type{list~\cic{term}} -> \boxty{\type{box\_type}} -> \type{list~name}\\
        & -> \type{result}~(\type{list~name} \times \boxty{\type{box\_type}})\\
        \ifdefined\ARXIV{}
          \erasetypeapp{&\Gamma_e}{\ident{\cic{args}}}{\ident{vs}}{\boxty{\sigma}} :=\\
          & \<match>~\ident{\cic{args}}~\<with>\\
        \else
          \erasetypeapp{\Gamma_e}{\ident{\cic{args}}}{\ident{vs}}{\boxty{\sigma}} :=&~ \<match>~\ident{\cic{args}}~\<with>\\
        \fi
        & ~|~ [] => \<Ok> (\ident{vs}, \boxty{\sigma})\\
        & ~|~ \cic{\ident{a}} ::\cic{\ident{args'}} =>\\
        & \qquad \cic{\ident{A}} <- \ident{type\_of}~\cic{\ident{a}};\\
        & \qquad \ident{flag} <- \flagoftype~\Gamma~\ident{\cic{A}};\\
        & \qquad \boxty{\tau} <- \<if>~(\<islogical>~\ident{flag})~\<then>~\<Ok>~\TBox\\
        & \qquad\qquad \<else>~\<if>~(\<issort> ~ \ident{flag})~\<then>\\
        & \qquad\qquad\quad(\ident{vs_\tau}, \boxty{\tau}) <- \erasetype{\Gamma}{\Gamma_e}{\ident{\cic{a}}}{\ident{vs}};\\
        & \qquad\qquad\quad\<if>~|\ident{vs}| < |vs_\tau|~\<then>~\error{\<NotPrenex>} \\
        & \qquad\qquad\quad \<else>~\<Ok>~\boxty{\tau}\\
        & \qquad\qquad \<else>~\<Ok>~\TAny;\\
        & \qquad \erasetypeapp{\Gamma_e}{\cic{\ident{args'}}}{\ident{vs}}
             {\boxty{\bigl({\sigma ~~ \tau}\bigr)}}\\
        & ~\<end>\\[.3em]
        \erasetypeheadname{} & : \type{ECtx} -> \type{\cic{term}} -> \type{result}~\boxty{\type{box\_type}}\\
        \ifdefined\ARXIV{}
          \erasetypeheadname&{\Gamma_e}{\ident{\cic{hd}}} :=\\
          & \<match>~\ident{\cic{hd}}~\<with>\\
        \else
          \erasetypeheadname{\Gamma_e}{\ident{\cic{hd}}} := &~ \<match>~\ident{\cic{hd}}~\<with>\\
        \fi
        & ~|~ \cic{\overline{i}} => \<match>~\Gamma_e(i)~\<with>\\
        & \qquad\quad |~\<RelInductive>~\ident{\cic{I}} => \ident{\boxty{I}}\\
        & \qquad\quad |~\_ => \error{\<Error>}\\
        & \qquad\quad \<end>\\
        & ~|~ \cic{C} => \<Ok>~\boxty{C}
        \quad|~ \cic{I} => \<Ok>~\boxty{I}
        \quad|~ \_ => \error{\<Error>}\\
        & ~\<end>\\[.3em]
        \erasevarname & : \type{ECtx} -> \mathbb{N} -> \boxty{box\_type}\\
        \ifdefined\ARXIV{}
          \erasevarname&~\Gamma_e~i :=\\
          & \<match>~\Gamma_e(i)~\<with>\\
        \else
          \erasevar{\Gamma_e}{i} := &~ \<match>~\Gamma_e(i)~\<with>\\
        \fi
        & ~|~\<RelTypeVar>~i => \boxty{\overline{i}}\quad |~\<RelOther> => \TBox{}
            \quad|~\<RelInductive>~\ident{\cic{I}} => \ident{\boxty{I}}\\
        & ~\<end>\\
      \end{aligned}
  \end{equation*}}\vspace{-10pt}
  \setlength{\belowcaptionskip}{-8pt}
  \caption{Erasure from \CIC{} types to \icode{box\_type}}\label{fig:type-erasure}
\end{figure*}
Currently, the verified erasure of MetaCoq provides only a term erasure procedure which will erase any type in a term to a box.
For example, a function using sigma types might have a signature involving \icode{sig nat (fun n => n > 10)}, i.e. representing numbers that are larger than $10$.
Applying MetaCoq's term erasure will erase this in its entirety to a box, while we are interested in a procedure that instead erases only the type scheme in the second argument: we expect type erasure to produce $\boxty{\type{sig} ~~ \type{nat} ~~ \square}$, where the square now represents an irrelevant type.

While our target languages have type systems that are Hindley-Milner based, for which type inference is normally regarded as complete, we still require an erasure procedure for types to be able to extract inductive types.
Moreover, our target languages support various extensions and their compilers may not always succeed to infer types; for example, Liquidity has overloading of some primitive operations (e.g.\ arithmetic operations for primitive numeric types) which introduces ambiguities that cannot be resolved by the type checker without type annotations.
Thus, the erasure procedure for types is also necessary to produce such type annotations.

The type systems in our target languages generally support prenex polymorphism, so we implement an erasure procedure that can erase prenex-polymorphic Coq types, giving a list of type parameters and erased type as a result.
The implementation of this procedure is inspired by~\cite{letouzey04}.
The outline of the procedure is given in \cref{fig:type-erasure}.
We have chosen a semi-formal presentation in order to guide the reader through the actual implementation and avoid cluttering with technicalities of Coq.
Additionally, we use colors to distinguish between the \textcolor{cicterms}{\CIC{} terms} and the target \textcolor{boxtyterms}{erased types}.

The \erasetypename{} function takes four parameters.
The first is a context $\type{Ctx}$ represented as a list of assumptions.
The second is an erasure context $\type{ECtx}$ represented as a sized list (vector) that follows the structure of $\type{Ctx}$; it contains either a translated type variable $\<RelTypeVar>$, information about an inductive type $\<RelInductive>$, or a marker for items in $\type{Ctx}$ that do not fit into the previous categories $\<RelOther>$.
The last two parameters represent terms of \CIC{} corresponding to types and a list of names for type variables used later for printing and for identifying non-prenex types.
We do not provide syntax and semantics of \CIC{}, for more information we refer the reader to Section 2 of~\cite{CertErasure}.
The function has a monadic type $\type{result}~(\type{list~name} \times \boxty{\type{box\_type}})$, which is essentially an extended error monad.
We use the standard \texttt{do}-notation to chain monadic computations.
The result of the computation is a tuple consisting of a list of type variables and a $\boxty{\type{box\_type}}$, the grammar for which is the following:
{\small\vspace{-0.25\baselineskip}\begin{align*}
  \boxty{\sigma},~\boxty{\tau} & ::= & \boxty{\overline{i}} ~|~ \boxty{I} ~|~ \boxty{C} ~|~ \boxty{\sigma~~\tau} ~|~\boxty{\sigma --> \tau} ~|~ \TBox ~|~\TAny
\end{align*}}\vspace{-1.25\baselineskip}

\noindent Here $\boxty{\overline{i}}$ represents indices of type variables and \boxty{I} and \boxty{C} range over names of inductive types and constants respectively.
Essentially, $\boxty{\type{box\_type}}$ represents types of an OCaml-like functional language extended with constructors $\TBox$ (``logical'' types) and $\TAny$ (types that are not representable in the target language).
In many cases both $\TBox$ and $\TAny$ can be removed from the extracted code by optimisations, although $\TAny$ might require type coercions in the target language.

The functions $\erasetypename{}$ and $\erasetypeappname{}$ are defined by mutual recursion.
The $\decomposeapp{}$ function returns the head of an application and a (possibly empty) list of arguments.
We use the notations $| \ident{xs} |$ to denote the length of $\ident{xs}$.
In our implementation, we extensively use dependently typed programming, so the actual type signature of functions in \cref{fig:type-erasure} contains also proofs that terms are well-typed.
The termination argument is given by a well-founded relation, since the erasure starts out with $\beta\iota\zeta$-reduction using the $\redbetaiotazeta$ function and then later recurses on subterms of this.
Here $\beta$ is reduction of applied $\lambda$-abstractions, $\iota$ is reduction of $\<match>$ on constructors, and $\zeta$ is reduction of the $\<let>$ construct.
The $\redbetaiotazeta$ function reduces until the head cannot be $\beta\iota\zeta$-reduced anymore and then stops; it does not recurse on subterms.
This reduction function is defined in MetaCoq also by using well-founded recursion.
Due to the well-founded recursion we write $\erasetypename{}$ as a single function in our formalization by inlining the definition of $\erasetypeappname{}$; this makes the well-foundedness argument easier.
We extensively use the \texttt{Equations} Coq plugin~\cite{Equations} in our development to help managing the proof obligations related to well-typed terms and recursion.

An important device used to determine erasable types (the ones we turn into the special target types $\TBox$ and $\TAny$) is the function $\flagoftype{}:~\type{Ctx} -> \type{\cic{term}} -> \type{type\_flag}$, where the return type $\type{type\_flag}$ is defined as a record with three projections: $\<islogical>$, $\<isarity>$ and $\<issort>$.\footnote{In our implementation, $\<islogical>$ carries a boolean, while  $\<isarity>$ and $\<issort>$ carry proofs or disproofs of convertibility to an arity or sort, respectively.}

A type is an arity if it is a (possibly nullary) product into a sort: $\forall \vec{\cic{a}} : \vec{\type{\cic{A}}}, \type{\cic{s}}$ for $\cic{\type{s}} = \cic{\type{Type}} ~|~ \cic{\type{Prop}}$ and $\vec{\cic{a}} : \vec{\type{\cic{A}}}$ a vector of (possibly dependent) binders and types. Inhabitants of arities are \textit{type schemes}.

$\<issort>$ tells us if a given type is a sort, i.e. \ $\cic{\type{Prop}}$ or $\cic{\type{Type}}$.
Sorts are always arities.
Finally, a type is logical when it is a proposition (i.e. inhabitants are proofs) or when it is an arity into $\type{\cic{Prop}}$: $\forall \vec{\cic{a}} : \vec{\type{\cic{A}}}, \type{\cic{Prop}}$ (i.e. inhabitants are propositional type schemes).

As concrete examples, $\cic{\type{Type}}$ is an arity and a sort, but not logical.
$\cic{\type{Type}} -> \cic{\type{Prop}}$ is logical, an arity, but not a sort.
$\cic{\<forall>~A : \type{Type},~A -> A}$ is neither of the three. Using erasure for types, we implement an erasure procedure for inductive definitions.

\paragraph{Optimisations}
Our second extension of the certified erasure is \emph{deboxing} --- a simple optimisation procedure for removing some redundant constructs (boxes) left after the erasure step.
First, we observe that removing redundant boxes is a special case of more general optimisation: elimination of dead arguments.
Informally it boils down to the equivalence $(\lambda x.~t)~v \sim t$ when $x$ does not occur free in $t$.
Here $\sim$ means that both sides evaluate to the same value.
Then, deboxing becomes a special case: $(\lambda A~x.~t)~\square \sim \lambda x.~t$.
From erasure, we know that the variable $A$ does not occur free in \icode{t}.\footnote{In our implementation we do not rely on this property and instead more generally remove unused parameters.}
Having in mind this equivalence, we implement in Coq a function with the following signature:
\vspace*{-1pt}
{\small\begin{align*}
  \dearg~:&~ \type{ind\_masks} -> \type{cst\_masks} -> \type{term} -> \type{term}
\end{align*}}%
The first two parameters are lookup tables for inductive definitions and for constants defining which arguments of constructors and constants to remove.
The type $\type{term}$ represents \CICbox{} terms.
The $\dearg$ function traverses the term and adjusts all applications of constants and constructors using the masks.

We define the following function that processes the definitions of constants:
{\small\begin{align*}
  \deargcst ~:&~ \type{ind\_masks} -> \type{cst\_masks} -> \type{constant\_body}\\
  & -> \type{constant\_body}
\end{align*}}%
This function deargs the body using $\dearg$ and additionally removes lambda abstractions in correspondence to the mask for the current constant.
Note that, since the masks apply only to constants in the program, we only remove dead parameters of top-level functions: abstractions representing closures are untouched.
Additionally, as dearging removes parameters from top level function, we must adjust the type signatures produced by the type erasure correspondingly.

To generate the masks we implement an analysis procedure that finds dead parameters of constants and dead constructor arguments.
For parameters of constants we check syntactically if they do not appear in the body, while for constructor arguments we find all unused arguments in pattern matches and projections across the whole program.
This is implemented as a linear pass over each function body that marks all uses of arguments and constructor arguments in that function.
As we noted above the erased arguments will be unused and therefore this procedure gives us a safe way of removing many redundant boxes (cf. Section 4.3 in~\cite{letouzey04}).

The syntactic check is quite imprecise; for example, it will not remove a parameter if its only use is to be passed to another function in which it is also unused.
To deal with this the analysis and dearging procedure can be iterated multiple times, but since our main use of the dearging is to remove arguments that are erased, this is not necessary.

For definitions of inductive types, we define the function
{\small\begin{align*}
  \deargmib ~ :~& \type{mib\_masks} -> \mathbf{N} -> \type{one\_inductive\_body} \\
  & -> \type{one\_inductive\_body}
\end{align*}}%
which adjusts the definition of one inductive's body of a (possibly) mutual inductive definition.
With $\deargcst$ and $\deargmib$, we can now define a function that removes arguments according to given masks for all definitions in the global environment:
{\small\begin{align*}
        \deargenv~:&~ \type{ind\_masks} -> \type{cst\_masks} -> \type{global\_env}\\ & -> \type{global\_env}
\end{align*}}%
Dearging is then done by first analyzing the environment to obtain $\type{ind\_masks}$ and $\type{cst\_masks}$ and then applying the $\deargenv$ function.

We prove dearging correct under several assumptions on the masks and the program being erased.

First, we assume that all definitions in the program are closed, which is a reasonable assumption given by typing.

Secondly, we assume validity of all the masks, meaning that all removed arguments of constants and constructors should be unused in the program.
By unused we mean that the argument does not syntactically appear except for in the binder.
The analysis phase outlined above is responsible for generating masks that satisfy this condition, although currently we do not prove this and instead recheck that the condition holds for the masks that were output.

Finally, we assume that the program is $\eta$-expanded according to all the masks: all occurrences of constructors and constants should be applied enough.
We implement a \emph{certifying} procedure that performs $\eta$-expansion and generates proofs that the expanded terms are equal to the original ones.
Since $\eta$-conversion is part of the Coq's conversion, the proofs are essentially just applications of the constructor \icode{eq_refl}.\footnote{See \texttt{extraction/examples/CounterDepCertifiedExtraction.v} for an example of using the technique in the extraction pipeline.}

Our Coq formalisation features a proof of the following soundness theorem about the $\dearg$ function.
\begin{theorem}[Soundness of dearging]\label{thm:dearg-sound}
    Let $\boxty{\Sigma}$ be a closed erased environment and $\boxty{t}$ a closed \CICbox{}-term such that $\boxty{\Sigma}$ and $\boxty{t}$ are valid and expanded according to provided masks.\\
    Then
    \vspace{-0.25\baselineskip}\[ \wcbveval{\boxty{\Sigma}}{\boxty{t}}{\boxty{v}} \]
    implies
    \vspace{-0.25\baselineskip}\[ \wcbveval{\deargenv(\boxty{\Sigma})}{\dearg(\boxty{t})}{\dearg(\boxty{v})} \]
    where dearging is done using the provided masks.
\end{theorem}%
\noindent
Here $\wcbveval{\boxty{-}}{\boxty{-}}{\boxty{-}}$ denotes the big-step call-by-value evaluation relation of \CICbox{} terms\footnote{The relation is part of MetaCoq. We contributed to fixing some issues with the specification of this relation.} and values are given as a subset of terms.
The theorem ensures that the dynamic behaviour is preserved by the optimisation function.
This result, combined with the fact that the erasure from \CIC{} to \CICbox{} preserves dynamic behaviour as well gives us guarantees that the terms that evaluate in \CIC{} will be evaluated to related results in \CICbox{} after optimisations.

\cref{thm:dearg-sound} is a relatively low level statement talking about the dearging optimisation that is used by our extraction.
The extraction pipeline itself is more complicated and works as outlined at the end of \cref{sec:smart-contract-extraction}: it is provided a list of definitions to extract in a well-typed environment and recursively erases these and their dependencies (see the full pipeline in \cref{fig:pipeline}).
Note that only dependencies that appear in the erased definitions are considered as dependencies; this typically gives an environment that is substantially smaller than the original.
Once the procedure has produced an environment, the environment is analysed to find out which arguments can be removed from constructors and constants, and finally the dearging procedure is invoked.

MetaCoq's correctness proof of erasure requires the full environment to be erased.
Since we only erase dependencies we prove a strengthened version of their theorem that is applicable for our case.
Combining this with \cref{thm:dearg-sound} allows us to obtain a statement about the full extraction pipeline (excluding pretty-printing).

\begin{theorem}[Soundness of extraction]\label{thm:extract-sound}
    Let $\cic{\Sigma}$ be a well-typed axiom-free environment and let $\cic{C}$ be a constant in $\cic{\Sigma}$.
    Let $\boxty{\Sigma'}$ be the environment produced by successful extraction (including optimisations) of $\cic{C}$ from $\cic{\Sigma}$.
    Then, for any unerasable constructor \cic{Ctor}, if
    \vspace{-0.25\baselineskip}\[ \wcbvevalpcuic{\cic{\Sigma}}{\cic{C}}{\cic{Ctor}} \]
    it holds that
    \vspace{-0.25\baselineskip}\[\wcbveval{\boxty{\Sigma'}}{\boxty{C}}{\boxty{Ctor}}\]
\end{theorem}%
\noindent
Here $\wcbvevalpcuic{\cic{-}}{\cic{-}}{\cic{-}}$ denotes the big-step call-by-value evaluation relation for CIC terms.
Informally, the above statement can be specialised to say that any program computing a boolean value will compute the same value after extraction.
Of course, one still has to keep in mind that the pretty-printing step of the extracted environment is not verified and the discrepancies of \CICbox{} and the target language's semantics as we outlined in \cref{sec:smart-contract-extraction}.

While the statement does not say anything about constructor applications,\footnote{It is hard to give an easily understandable statement since dearging removes applications.} it does informally generalise to any value that can be encoded as a number, since it can be used to show that each bit of the output will be the same.

One of the premises of \cref{thm:extract-sound} is that the environment is axiom-free, which is required for the soundness of erasure as stated in MetaCoq and adapted in our work.
In general, we cannot say anything about the evaluation of terms once axioms are involved.
One possible way of fixing this issue is by following the semantic approach as in Section 2.4 of~\cite{letouzey04}.

While dearging subsumes deboxing we cannot guarantee that our optimisation removes all boxes even for constants applied to all logical arguments due to cumulativity.
\footnote{By cumulativity we mean subtyping for universes, i.e. \icode{A : Type}$_i$ is also \icode{A : Type}$_{i + 1}$ for any $i$.
Therefore, if a function takes an argument \icode{A : Type}, we can pass \icode{Prop}, since it is at the lowest level of the universe hierarchy.}
E.g.\ for \icode{@inl Prop Prop True : sum Prop Prop} it is tempting to optimise the extracted version \icode{inl ∎ ∎ ∎} into just \icode{inl}, but the optimised definition of the \icode{sum} type will still have the \icode{inl} constructor that takes one argument, because its type is \icode{inl : forall A B : Type, A -> A + B} and the argument \icode{A} is in general relevant for computations.

As mentioned previously, the dearging of functions removes parameters which means that it must also adjust the type signatures of those functions.
In addition to this adjustment of type signatures, we also do a final pass to remove logical inductive type parameters.
This step is completely orthogonal to the dearging of terms and serves only to remove useless type parameters.
This does not affect the dynamic semantics, but mistakes in it might mean that the code does not type check in the target language.

For a concrete example, sigma types are defined in Coq as
\begin{lstlisting}
Inductive sig (A : Type) (P : A -> Prop) :=
  exist : forall x : A, P x -> sig A P
\end{lstlisting}%
In the constructor, \icode{P} is a type scheme while the argument of type \icode{P x} is a proof, so these are erased by type erasure, resulting in the type $\boxty{\type{A} --> \square --> \type{sig} ~~ \type{A} ~~ \square}$.
The analysis will show that the proof argument is never used, since any use is also erased.
This means the constructor is changed to $\boxty{\type{A} --> \type{sig} ~~ \type{A} ~~ \square}$ as part of the dearging process, and any use of this constructor in a function (e.g. for pattern matching, or to construct a value) is similarly adjusted.
Finally, removal of logical type parameters means that the type parameter $P$ is completely removed from the type, giving the final constructor type as $\boxty{\type{A} --> \type{sig} ~~ \type{A}}$.
Function signatures using \icode{sig} are also adjusted correspondingly, having the $P$ argument removed.

After applying the optimisations we pretty-print the optimised code to a functional smart contract language.
We use the fact that \CICbox{} is similar to a subset of a generic functional language.
Therefore, a pretty-printer can take an extracted and deboxed term and produce a program in Liquidity, Midlang or another language that is sufficiently close to \CICbox{}.
Currently, we support most of the constructs of \CICbox{} in our code generators, apart from primitive projections and coinductive definitions.
We discuss issues related to extraction to Liquidity in \cref{sec:liquidity-extraction} and to Midlang in \cref{sec:midlang-extraction}.

\begin{figure}
  \begin{center}
    \begin{minipage}{0.6\textwidth}
      \begin{lstlisting}
Definition storage := Z.
Inductive msg := Inc (_ : Z) | Dec (_ : Z).
Program Definition inc_counter (st : storage) (inc : {z : Z | 0 <? z}) :
  {new_st : storage | st <? new_st} := st $+$ inc. (* proof omitted *)
Program Definition dec_counter (st : storage) (dec : {z : Z | 0 <? z}) :
  {new_st : storage | new_st <? st} := st $-$ dec. (* proof omitted *)
Definition my_bool_dec := Eval compute in bool_dec.

Definition counter (msg : msg) (st : storage)
  : option (list operation * storage) :=
  match msg with
  | Inc i => match (my_bool_dec (0 <? i) true) with
    | left h => Some ([], proj1_sig (inc_counter st (exist _ i h)))
    | right _ => None
    end
  | Dec i => match (my_bool_dec (0 <? i) true) with
    | left h => Some ([], proj1_sig (dec_counter st (exist _ i h)))
    | right _ => None
    end
  end.
      \end{lstlisting}
    \end{minipage}
  \end{center}
\vspace{-15pt}
\setlength{\belowcaptionskip}{-8pt}
\caption{The \texttt{counter} contract}\label{lst:counter}
\end{figure}

\paragraph{The Counter Contract}
As an example, let us consider a simple counter contract with the state being just an integer number and accepting increment and decrement messages (\cref{lst:counter}).
The main functionality is given by the two functions \icode{inc_counter} and \icode{dec_counter}. We use \emph{refinement types} to encode the functional specification of these functions.
E.g.\ for \icode{inc_counter} we encode in the type that the result of the increment is greater than the previous state given a positive increment.
Refinement types are represented in Coq as dependent pairs ($\Sigma$-types).
For example a positive integer is encoded as \icode{\{z : Z | 0 <? z\}}, where the second component is a proposition \icode{0 <? i = true} (we use an implicit coercion from booleans to propositions).
Similarly, we encode the specification \icode{dec_counter}.
The \icode{counter} function validates the input and provides a proof that the input satisfies the precondition (of being positive).
The functions \icode{inc_counter} and \icode{dec_counter} are defined only for positive increments and decrements, therefore, we do not need to validate the input again.
Note that in order to construct an inhabitant of \icode{positive} we use the decidability of equality for booleans \icode{bool_dec : forall b1 b2 : bool, \{b1 = b2\} + \{b1 <> b2\}}\footnote{We simplify \icode{bool_dec} using \icode{Eval compute in ...} in order to unfold the recursor \icode{bool_rec} in the definition of \icode{bool_dec}. The types of recursors are not in the prenex form, therefore they give an error during erasure. Eventually, we want to implement a separate pass for unfolding constants like \icode{bool_rec}.} that gives us access to the proof of \icode{0 <? i}.
We will use the example from \cref{lst:counter} in subsequent sections for showing how it can be extracted to concrete target languages.

\lstset{basicstyle=\scriptsize}
\begin{figure*}
  \begin{subfigure}[t]{.47\textwidth}
    \begin{lstlisting}
type 'a sig_ = 'a
let exist_ a = a
type coq_msg = Coq_Inc of int | Coq_Dec of int
type storage = int
type coq_sumbool = Coq_left | Coq_right

let coq_inc_counter (st : storage) (inc : int sig_) =
     exist_ (addInt st ((fun x -> x) inc))
     ...

let coq_counter (msg : coq_msg) (st : storage) =
match msg with
  Coq_Inc i ->
  (match coq_my_bool_dec (ltInt 0 i) true with
    Coq_left  ->
    Some ([], ((fun x -> x)
       (coq_inc_counter st (exist_ (i)))))
  | Coq_right  -> None)
| Coq_Dec i -> ...
  | Coq_right  -> None)
    \end{lstlisting}
    \vspace{-5pt}
    \caption{Liquidity}\label{fig:extracted-liquiidty}
  \end{subfigure}
  \begin{subfigure}[t]{.47\textwidth}
    \begin{lstlisting}
type Sig a  = Exist a
type Msg = Inc Int | Dec Int
type alias Storage = Int
type Sumbool = Left | Right

proj1_sig : Sig a -> a
proj1_sig e = case e of
    Exist a -> a

inc_counter : Storage -> Sig Int -> Sig Storage
inc_counter st inc = Exist (add st (proj1_sig inc))
...
counter : Msg -> Storage -> Option (Prod Transaction Storage)
counter msg st =
  case msg of
    Inc i -> case my_bool_dec (lt 0 i) True of
              Left -> Some (Pair Transaction.none (proj1_sig (inc_counter st (Exist i))))
              Right -> None
    Dec i -> ...
    \end{lstlisting}
    \vspace{-5pt}
    \caption{Midlang}\label{fig:extracted-midlang}
  \end{subfigure}
  \setlength{\abovecaptionskip}{5pt}
  \setlength{\belowcaptionskip}{-8pt}
  \caption{Extracted code.}\label{fig:extracted-code}
\end{figure*}

\lstset{basicstyle=\footnotesize}

\subsection{Extracting to Liquidity}\label{sec:liquidity-extraction}
Liquidity is a functional smart contract language for the Tezos and Dune blockchains inspired by OCaml.
It compiles to Michelson developed by Tezos --- a stack-based functional language that runs directly on the blockchain.
Compared to a conventional functional language, Liquidity has many restrictions.
E.g.\ data types are limited to non-recursive inductive types, support for recursive definitions is limited to tail recursion on a single argument.
That means that one is forced to use primitive container types to write programs.
Therefore, the functions on lists and finite maps must be replaced with ``native'' versions in the extracted code.
We achieve this by providing a translation table that maps names of Coq functions to the corresponding Liquidity primitives.
Moreover, since the recursive functions can take only a single argument, multiple arguments need to be packed into a tuple.
The same applies to data type constructors since the constructors take a tuple of arguments.
Currently, the packing into tuples is done by the pretty-printer after verifying that constructors are fully applied.

Another issue is related to the type inference in Liquidity.
Due to the support of overloaded operations on numbers, type inference is requires type annotations.
We solve this issue by providing a ``prelude'' for extracted contracts that specifies all required operations on numbers with explicit type annotations.
This also simplifies the remapping of Coq operations to the Liquidity primitives.
Moreover, we produce type annotations for top-level definitions.

In order to generate code for a contract's entry points (functions through which one can interact with the contract) we need to wrap the calls to the main functionality of the contract into a \icode{match} construction.
This is required because the signature of the entry point in Liquidity is \icode{params -> storage -> (operation list)$\ $* storage }, where \icode{params} is a user-defined type of parameters, \icode{storage} a user-defined state and \icode{operation} is a transfer of contract call.
The signature looks like a total function, but since Liquidity supports a side effect \icode{failwith}, the entry-point function still can fail.
On the other hand, in our Coq development, we use the \icode{option} monad to represent computations that can fail.
For that reason, we generate a wrapper that matches on the result of the extracted function and calls \icode{failwith} if it returns \icode{None}

The extracted counter contract code is given in \cref{fig:extracted-liquiidty}.
We omit some wrapper code and the ``prelude'' definitions and leave the most relevant parts (see \cref{appendix:counter-liquidity} for the full version).
As one can see, the extraction procedure removes all ``logical'' parts from the original Coq code.
Particularly, the \icode{sig} type of Coq becomes a simple wrapper for a value (\icode{type 'a sig_ = 'a} in the extracted code).
Currently, we resort to an ad hoc remapping of \icode{sig} to the wrapper \icode{sig_} because Liquidity does not support variant types with only a single constructor.
Ideally, this class of transformations can be added as an optimisation for inductives types with only one constructor taking a single argument.
This example shows that for certain target languages optimisation is a necessity rather than an option.

We show the extracted code for the \icode{coq_inc_counter} function and omit \icode{coq_dec_counter}, which is extracted in a similar manner.
These functions are called from the \icode{counter} function that performs input validation.
Since the only way of interacting with the contract is by calling \icode{counter} it is safe to execute them without additional input validation, exactly as it is specified in the original Coq code.

Apart from the example in \cref{lst:counter}, we successfully applied the developed extraction to several variants of the counter contract, to the crowdfunding contract described in~\cite{ConCert} and to an interpreter for a simple expression language. The latter example shows the possibility of extracting certified interpreters for domain-specific languages such as Marlowe~\cite{Marlowe}, CSL~\cite{Henglein:CSL} and the CL language~\cite{CertFinContr,CertifiedCompCL}.
This represents an important step towards safe smart contract programming.
The examples show that smart contracts fit well to the fragment of Coq that extracts to well-typed Liquidity programs.
Moreover, in many cases, our optimisation procedure removes all the boxes resulting in cleaner code.

\vspace*{-2mm}\subsection{Extracting to Midlang and Elm}\label{sec:midlang-extraction}
Midlang is a functional smart contract language for the Concordium blockchain.
It is a fork of Elm~\cite{ElmInAction} --- a general-purpose functional language used for web development.
Being close to Elm means that it is a fully-featured functional language that supports many usual functional programming idioms.
Compared to Liquidity, Midlang is a better target for code extraction, since it does not have the limitations pointed out in \cref{sec:liquidity-extraction}.

We use the same example from \cref{lst:counter} to demonstrate extraction to Midlang (\cref{fig:extracted-midlang}, see \cref{appendix:counter-midlang} for the full version).
Similarly to Liquidity, extracted code does not contain logical parts (e.g.\ proofs of being positive).
The \icode{sig} type of Coq extracts to the type definition \icode{Sig} with a single constructor being a simple wrapper around the value.
In Midlang we do not have to ``unwrap'' the value from the \icode{Exist} constructor in an ad-hoc way since single constructor data types are allowed, but one could still imagine this as an optimisation.

Extraction to Midlang poses some challenges which are inherited from Elm.
For example, Midlang does not allow shadowing of variables and definitions.
Since Coq allows for a more flexible approach to naming, one has to track scopes of variables and generate fresh names in order to avoid clashes.
The syntax of Midlang is indentation sensitive that requires tracking of indentation levels.
Various naming conventions apply to Midlang identifiers, e.g.\ function names start with a lower-case character, types and constructors --- with an upper case character, requiring some names to be changed when printing.

We have tested the support for Midlang extraction on several examples including the contract from \cref{lst:counter} and the escrow contract that we will describe in \cref{sec:escrow}.
Both the counter and the escrow contracts were successfully extracted and compiled with the Midlang compiler.
However, the escrow contract requires more infrastructure for mapping the ConCert blockchain formalisation definitions to the corresponding Midlang primitives.
Since Midlang is a fork of Elm~\cite{ElmInAction}, code that does not use any blockchain specific primitives is also extractible to Elm.
We tested the extracted code with Elm compiler by generating a simple test for each extracted function.
We implemented several tests extracting functions on lists from Coq's standard library and functions using refinement types.
The \icode{safe_head} (a head of a non-empty list) example uses the elimination principle \icode{False_rect}.
We support this by an ad hoc remapping of \icode{False_rect} to \icode{false_rec _ = false_rec ()}.
Since we know that the impossible case never happens, we can use this ``infinite loop'' function in its place.
Moreover, we extracted the Ackermann function \icode{ackermann : nat * nat -> nat} defined using well-founded recursion which uses the lexicographic ordering on pairs.
This shows that extraction of definitions based on the accessibility predicate \icode{Acc} is possible.
Computation with \icode{Acc} is studied in more detail in~\cite{Equations}.


\section{The Escrow Contract}\label{sec:escrow}
As an example of a nontrivial contract we can extract we describe in this section an \textit{escrow} contract.\footnote{See \texttt{execution/examples/Escrow.v} in the artifact.}
The purpose of this contract is to enable a seller to sell goods in a trustless setting via the blockchain.
The Escrow contract is suited for goods that cannot be delivered digitally over the blockchain; for goods that can be delivered digitally, there are contracts with better properties, such as FairSwap~\cite{FairSwap}.

Because goods are not delivered on chain there is no way for the contract to verify that the buyer has received the item.
Instead, we incentivise the parties to follow the protocol by requiring that both parties commit additional money that they are paid back at the end.  
Assuming a seller wants to sell a physical item for $x$ amount of currency, the contract proceeds in the following steps:
\begin{enumerate}
\item The seller deploys the contract and commits (by including with the deployment) $2x$.
\item The buyer commits $2x$ before a deadline.
\item The seller delivers the goods (outside of the smart contract).
\item The buyer confirms (by sending a message to the smart contract) that he has received the item.
      He can then withdraw $x$ from the contract while the seller can withdraw $3x$ from the contract.
\end{enumerate}

If there is no buyer who commits funds the seller can withdraw his money back after the deadline.
Note that when the buyer has received the item, he can choose not to notify the smart contract that this has happened.
In this case he will lose out on $x$, but the seller will lose out on $3x$.
In our work we assume that this does not happen, and we consider the exact game-theoretic analysis of the protocol to be out of scope.
Instead, we focus on proving the \textit{logic} of the smart contract correct under the assumption that both parties follow the protocol to completion.
The logic of the Escrow is implemented in around a hundred lines of Gallina code.
The interface to the Escrow is its message type given below.
\begin{lstlisting}
Inductive Msg := commit_money | confirm_item_received | withdraw.
\end{lstlisting}
To state correctness we first need a definition of what the escrow's effect on a party's balance has been.
\begin{definition}[Net balance effect]
Let $\pi$ be an execution trace and $a$ be an address of some party.
Let $T_\text{from}$ be the set of transactions from the Escrow to $a$ in $\pi$, and let $T_\text{to}$ be the set of transactions from $a$ to the contract in $\pi$.
Then the net balance effect of the Escrow on $a$ is defined to be the sum of amounts in $T_\text{from}$, minus the sum of amounts in $T_\text{to}$.
\end{definition}
\noindent The Escrow keeps track of when both the buyer and seller have withdrawn their money, after which it marks the sale as completed.
This is what we use to state correctness.
\begin{theorem}[Escrow correctness]\label{thm:escrow-correct}
  Let $\pi$ be an execution trace with a finished Escrow for an item of value $x$.
  Let $S$ be the address of the seller and $B$ the address of the buyer.
  Then:
  \begin{itemize}
    \item If $B$ sent a \icode{confirm_item_received} message to the Escrow, the net balance effect on the buyer is $-x$ and the net balance effect on the seller is $x$.
    \item Otherwise, the net balance effects on the buyer and seller are both $0$.
  \end{itemize}
\end{theorem}
\noindent In \cref{sec:testing-escrow} we describe how this property can also be tested automatically by using QuickChick.

As mentioned earlier we have used our extraction to produce a Midlang version of the verified escrow contract, which gives us high guarantees with some caveats.
\cref{thm:escrow-correct} relies on the execution \emph{model} of the actual implementation of the blockchain.
Therefore, the execution model is part of the TCB.
However, the model provides a good approximation of execution layers used for \emph{functional} smart contracts.
While we have no formal proof that \cref{thm:escrow-correct} translates to the version produced by extraction, this still gives us a high level of confidence due to the certified erasure and optimisations.

\section{The Boardroom Voting Contract}\label{sec:boardroom}
Hao, Ryan and Zieli\'{n}sky developed the Open Vote Network protocol~\cite{open-vote-network}, an e-voting protocol that allows a small number of parties (`a boardroom') to vote anonymously on a topic.
Their protocol allows tallying the vote while still maintaining maximum voter privacy, meaning that each vote is kept private unless all other parties collude.
Each party proves in zero-knowledge to all other parties that they are following the protocol correctly and that their votes are well-formed.

This protocol was implemented as an Ethereum smart contract by McCorry, Shahandashti and Hao~\cite{ethereum-boardroom}.
In their implementation, the smart contract serves as the orchestrator of the vote by verifying the zero-knowledge proofs and computing the final tally.

We implement a similar contract in the ConCert framework.\footnote{See \texttt{execution/examples/BoardroomVoting.v} in the artifact.}
The original protocol works in three steps.
First, there is a sign up step where each party submits a public key and a zero-knowledge proof that they know the corresponding private key.
After this, each party publishes a commitment to their upcoming vote.
Finally, each party submits a computation representing their vote, but from which it is computationally intractable to obtain their actual private vote.
Together with the vote, they also submit a zero-knowledge proof that this value is well-formed, i.e.\ it was computed from their private key and a private vote (either `for` or `against`).
After all parties have submitted their public votes, the contract is able to tally the final result.
For more details, see the original paper~\cite{open-vote-network}.

The contract accepts messages given by the type:
\begin{lstlisting}
Inductive Msg :=
| signup (pk : A) (proof : A * Z)
| commit_to_vote (hash : positive)
| submit_vote (v : A) (proof : VoteProof)
| tally_votes.
\end{lstlisting}%
Here, \icode{A} is an element in an arbitrary finite field, \icode{Z} is the type of integers and \icode{positive} can be viewed as the type of finite bit strings.
Since the tallying and the zero-knowledge proofs are based on finite field arithmetic we develop some required theory about $\mathbb{Z}_p$ including Fermat's theorem and the extended Euclidean algorithm.
This allows us to instantiate the boardroom voting contract with $\mathbb{Z}_p$ and test it inside Coq using ConCert's executable specification.
To make this efficient, we use the Bignums library of Coq to implement operations inside $\mathbb{Z}_p$ efficiently.

The contract provides three functions \icode{make_signup_msg}, \icode{make_commit_msg} and \icode{make_vote_msg} meant to be used off-chain by each party to create the messages that should be sent to the contract.
As input these functions take the party's private data, such as private keys and the private vote, and produces a message containing derived keys and derived votes that can be made public, and also zero-knowledge proofs about these.
We prove the zero-knowledge proofs attached will be verified correctly by the contract when these functions are used.
Note that, due to this verification done by the contract, the contract is able to detect if a party misbehaves.
However, we do not prove formally that incorrect proofs do not verify since this is a probabilistic statement better suited for tools like EasyCrypt.

When creating a vote message using \icode{make_vote_msg} the function is given as input the private vote: either `for`, represented as $1$, and `against`, represented as $0$.
We prove that the contract tallies the vote correctly assuming that the functions provided by the boardroom voting contract are used.
Note that the contract accepts the \icode{tally_votes} message only when it has received votes from all public parties, and as a result stores the computed tally in its internal state.
We give here a simplified version of the full correctness statement which can be found in the attached artifact.
\begin{theorem}[Boardroom voting correct]
  Let $\pi$ be an execution trace with a boardroom voting contract.
  Assume that all messages to the Boardroom Voting contract in $\pi$ were created using the functions described above. Then:
  \begin{itemize}
    \item\sloppy If the boardroom voting contract has accepted a \icode{tally_votes} message, the tally stored by the contract equals the sum of private votes.
    \item Otherwise, no tally is stored by the contract.
  \end{itemize}
\end{theorem}
\noindent The boardroom voting contract gives a good benchmark for our extraction as it relies on some expensive computations.
It drives our efforts to cover more practical cases, and we are currently working on extracting it in a performant manner.

The main problem with extraction for this contract is the use of higher-kinded types.
In particular, the implementation of the contract uses finite maps from the std++ library, which implicitly rely on higher-kinded types.
In addition, the contract uses monadic binds, implemented via type classes which require passing type families around.
This is not representable in prenex-polymorphic type systems, and our target languages follow similar typing discipline.
While we could adjust the implementation to avoid relying on higher kinded types, we instead prefer to improve the extraction to work on more examples.
In particular, for our cases we have identified that a few steps of reduction is enough for the higher kinded types to disappear.
For example, the signature of \icode{bind} is \icode{forall m : Type -> Type, Monad m -> forall t u : Type, m t -> (t -> m u) -> m u} which, when it appears in the contract, typically looks like \icode{bind option option_monad ...} where \icode{option_monad} is some constant that builds a record describing the option monad.
After very few steps of reduction, this reduces to the well-known bind for options, which is unproblematic to extract.
We thus plan to resolve these problems and be able to extract even more examples by implementing a verified pass that unfolds and reduces certain constants before extraction.
In part, this also makes our extraction more like Coq's built-in extraction which also performs inlining (in an unverified manner).

\section{Property-Based Testing of Smart Contracts}\label{sec:testing}
With ConCert's executable specification our contracts are fully testable from within Coq.\footnote{The testing framework for smart contracts is available in the \texttt{execution/tests} subfolder of the artifact.}
This enables us to integrate property-based testing into ConCert using QuickChick~\cite{Paraskevopoulou:QuickChick}. 
It serves as a cost-effective, semi-automated approach to discover bugs and it increases reliability that the implementation is correct. 
Furthermore, since QuickChick is formally verified, we know that a reported counterexample is in fact a true negative of the property under test.
Testing may be used either as a preliminary step to support formal verification or as a complementary approach whenever the properties become too involved to prove.

Property-based testing is an automated, generative approach to software testing where test data is automatically generated and tested against some executable specification, and any failed test case is reported. 
As opposed to example-based testing, where the user manually constructs and executes a few test cases, property-based testing can cover a much larger input scope by generating thousands of ``arbitrary'' test data, which increases the potential of discovering bugs.

QuickChick is a property-based testing framework in Coq inspired by Haskell's Quick\emph{Check}~\cite{claessen:QuickCheck}. 
The user provides an executable specification (e.g.\ a decidable property) and input generators for the necessary data types. QuickChick will then test that all generated test cases pass, reporting any discovered counterexample. 
In many cases the input generators can be derived either partially or fully automatically, and QuickChick provides generators for common data types such as \icode{nat}, \icode{Z}, \icode{bool}, and \icode{list}. 
As a simple example, \cref{lst:qc-ex} shows how to test an inverse property between \icode{square} and \icode{sqrt} using QuickChick. 
QuickChick can execute this test because it has a built-in generator for arbitrary \icode{nat}s, and because equality on \icode{nat} is decidable, and therefore the entire property is decidable. 
Internally, QuickChick converts \icode{example\_prop} to the executable term \icode{y = square x ==> sqrt y = x} where \icode{==>} is an executable variant of implication that discards a test whenever the pre-condition is false.\footnote{In the example in \cref{lst:qc-ex} QuickChick reported 0 discards. 
This is because QuickChick is able to automatically derive a generator satisfying the inductive predicate \icode{y = square x}. This is in general not always possible.}

\begin{figure}
  \begin{center}
    \begin{minipage}{0.5\textwidth}
      \begin{lstlisting}
Conjecture example_prop :
  forall (x y : nat), y = square x -> sqrt y = x.
QuickChick example_prop.
(* Passed 10000 tests (0 discards) *)
      \end{lstlisting}
    \end{minipage}
  \end{center}
\vspace{-15pt}
\setlength{\belowcaptionskip}{-10pt}
\caption{Simple example usage of QuickChick.}\label{lst:qc-ex}
\end{figure}

Since the testing is intended to support verification, we should be able to test the same properties as those we wish to prove (assuming the property is decidable). 
These properties are usually stated in terms of blockchain execution traces. 
This poses the key question of how to generate ``arbitrary'' execution traces. 
An execution trace in ConCert is a sequence of blocks, each containing some number of \icode{Action}s (which may be either transactions, contract calls, or contract deployment).
Specifically, we must consider how to generate arbitrary contract calls for a given contract. 
Previous works on property-based testing of smart contracts such as Echidna~\cite{grieco:Echidna} and Brownie\footnote{Property-based testing framework for EVM:\url{https://github.com/eth-brownie/brownie}} employ a fuzzing-like approach where payload data of contract calls are populated with entirely random data. 
The advantage of this approach is that it can be completely automated, however the generated data may not provide good enough coverage for contracts with endpoints requiring complex conditions to be called.
In these cases, large proportions of the generated data will be discarded during testing, which leads to worse performance, lower test coverage, and worse expected bug discovering capability. 
Echidna mitigates this by using a coverage-based, self-improving algorithm for test generation.

We make a trade-off to overcome this issue by sacrificing some automation and instead require the user to supply specialised generators for the message type of the contracts under test, rather than automatically deriving these generators. 
From this, the framework automatically derives a generator of provably valid execution traces which is used for subsequent tests. 
For example, if the user supplies a generator for the \icode{Msg} type of the Escrow contract presented in \cref{sec:escrow}, the generated traces will contain contract calls to the Escrow contract (assuming it is deployed on the test chain) using this generator.

Our testing framework supports three kinds of testable properties: (1) a testable notion of universal quantification (which is realised by just executing, and asserting success of, 10.000 test cases) on any executable property on the \icode{ChainBuilder} type. 
This type represents a chain state along with a proof-relevant execution trace that led to this state. 
(2) A hoare-triple style pre- and post-condition assertions on the \icode{receive} function of a given contract, and (3) reachability of chain states satisfying some decidable property. Our development is the first to support testing on entire execution traces. This allows for testing properties of interacting smart contracts.
\begin{center}
  \vspace{-4pt}
  \begin{enumerate}
    \item \icode{forall (c : ChainBuilder), P c}
    \item \icode{\{pre\}SomeContract.receive\{post\}}
    \item \icode{init_chain $\ \leadsto\ $ (fun cs : ChainState => P cs)}
  \end{enumerate}
  \vspace{-4pt}
\end{center}

\subsection{Testing the Escrow Contract}\label{sec:testing-escrow}
The Escrow contract was described and proved correct in \cref{sec:escrow}. The entire Coq proof, including auxiliary lemmas about the Escrow contract, is about 500 lines.
Thus, it is a significant effort and in the presence of bugs in the implementation much time and effort could be wasted.
We demonstrate how to use the testing framework as a preliminary step in formal verification to potentially discover any bugs. 
The correctness property of the Escrow was defined in \cref{thm:escrow-correct}. 
Supposing the Escrow is finished in some block, then calculating the net balance effect is a decidable proposition, and can easily be proven so by coercing it to a \icode{bool}.
In order to be able to test this property we first need to derive a generator for traces containing calls to a deployed Escrow contract.
In \cref{lst:gEscrowMsg} we show such a generator.

\begin{figure}
  \begin{center}
    \begin{minipage}{0.6\textwidth}
      \begin{lstlisting}
Context {contract_addr : Address}.
Definition gEscrowMsg (e : Environment) (state : Escrow.State)
  : G (Option Action) :=
  (* creates a call to the escrow contract with some msg *)
 let call caller amount msg := ret (build_act caller
    (act_call contract_addr amount (serialize msg))) in
 (* pick one gen at random, backtrack if it fails *)
 backtrack [(1, if e.(account_balance) state.(buyer) <? 2
                 then returnGen None
                 else call state.(buyer) 2 commit_money) ;
              (1, call state.(buyer) 0 confirm_item_received) ;
              (1, addr <- elems [state.(seller); state.(buyer)] ;;
                  call addr 0 withdraw) ].  
      \end{lstlisting}
    \end{minipage}
  \end{center}
  \vspace{-10pt}
  \setlength{\belowcaptionskip}{-15pt}
  \caption{Generator of ``arbitrary'' Escrow messages.}\label{lst:gEscrowMsg}
  \end{figure}
\noindent
Omitting some boilerplate code for the initial chain setup, we can now test the correctness property with:
\begin{lstlisting}
  QuickChick (forall c : ChainBuilder, escrow_correct_P c).
  (* Passed 10000 tests (23743 discards) *)
\end{lstlisting}
If we manually insert a bug where the withdrawable amount is calculated differently, then QuickChick immediately reports a counterexample.
Aside from the example shown here, we have tested various other contracts implemented in ConCert such as ERC-20 tokens, FA2 tokens, a congress contract similar to ``The DAO'', and a liquidity exchange protocol consisting of multiple interacting contracts.
We used the testing framework to successfully discover well known contract vulnerabilities:
\begin{itemize}
  \item The ``DAO attack'' where, at the time, \$50 million worth of cryptocurrency was stolen due to a re-entrancy bug.\footnote{\url{https://www.wired.com/2016/06/50-million-hack-just-showed-dao-human/}}
  \item An exploit of the UniSwap exchange protocol on the Lendf.me platform where an attacker could perform a re-entrancy attack during token exchange to obtain arbitrary profit.\footnote{\url{https://forum.openzeppelin.com/t/exploiting-uniswap-from-reentrancy-to-actual-profit/1116}}
  This attack was possible for any ERC-777 compliant tokens.
  \item A recent exploit of the iToken contract where an attacker could mint arbitrary tokens for themselves by performing a self-transfer due to a bug in the implementation of the transfer function.\footnote{\url{https://bzx.network/blog/incident}}
\end{itemize}
This shows the effectiveness of our approach, not just for testing contracts in isolation, but also for testing multiple, interacting contracts.

\section{Related Work}\label{sec:related-work}
Works related to the extraction part can be split in two categories.
The most relevant related works are concerned with extraction to statically typed functional programming languages.
Several proof assistants share this feature (Coq~\cite{CoqNewExtraction}, Isabelle~\cite{Isabelle-ExecutingHOL}, Agda~\cite{Kusee:agda-haskell}) and allow targeting conventional functional languages such as Haskell, OCaml or Standard ML.
However, extraction in Isabelle/HOL~\cite{Isabelle-ExecutingHOL} is slightly different from Coq and Agda, since the in higher-order logic of Isabelle/HOL programs are represented as equations and the job of the extraction mechanism is to turn them into executable programs.
Clearly, the correctness of the extraction code is crucial for producing correct executable programs.
This is addressed by several developments for Isabelle~\cite{Isabelle-CodeGen,VerfiedCompIsabelle}.
The work~\cite{VerfiedCompIsabelle} features verified compilation from Isabelle/HOL to CakeML~\cite{CakeML}.
It also implements meta-programming facilities for quoting Isabelle/HOL terms similar to MetaCoq.
Moreover, the quoting procedure produces a proof that the quoted terms corresponds to the original ones.
The current extraction implemented in the Coq proof assistant is not verified, however, the theoretical basis for it is well-developed by Letouzey~\cite{letouzey04}.
On the other hand, Coq's extraction also includes unverified optimisations that are done together with extraction, making it harder to compare it with the formal treatment given by Letouzey. So, the unverified extraction even lacks a full paper proof.
Our separation between erasure and optimisation facilitates such comparisons, and allows reuse of the optimisation pass in a standalone fashion in other projects.
The MetaCoq project~\cite{CertErasure} aims to formalise the meta-theory of the Calculus of Inductive Constructions and features a verified erasure procedure that forms the basis for extraction presented in this work.
We also emphasise that the previous works on extraction targeted conventional functional languages (e.g.\ Haskell, OCaml, etc.), while we target the more diverse field of functional smart contract languages.

Another category of related approaches focuses on execution of dependently typed languages.
Although the techniques used in these approaches are similar, one does not need to fit the extracted code into the type system of a target language.
The dependently typed programming language Idris uses erasure techniques for efficient execution~\cite{Ind-falimilies-indices:Brady}.
A master's thesis~\cite{IdrisSmarContracts:Pettersson2016} explores the applicability of dependent types to smart contract development and extends the Idris compiler with Ethereum Virtual Machine code generation.
For the Coq proof assistant, the work~\cite{CIC-type-decorations:Barras2005} develops an approach for efficient convertibility testing of untyped terms acquired from fully typed CIC terms.
The {\OE}uf project~\cite{Oeuf} features verified compilation of a restricted subset of Coq's functional language Gallina (no pattern-matching, no user defined inductive types --- only eliminators for particular inductives).
In~\cite{ExtractionFiat}, the authors report on extraction of embedded into Gallina domain-specific languages into an imperative intermediate language which can be compiled to efficient low-level code.
And finally, the certified compilation approach to executing Coq programs is under development in the CertiCoq project~\cite{Anand:CertiCoq}.
The project uses MetaCoq for quotation functionality and uses the verified erasure as the first stage.
After several intermediate stages, C light code is emitted and later compiled for a target machine using the CompCert certified compiler~\cite{2006-Leroy-compcert}.
Since we implement our pass as a standalone optimisation on the same AST that is used in CertiCoq, our pass can be integrated in a relatively straightforward fashion in CertiCoq.
We are currently working with the authors of CertiCoq on making this integration.


The boardroom voting is based on the Open Vote Network by Hao, Ryan and Zieli\'{n}sky~\cite{open-vote-network}.
In their paper there are paper proofs showing the computation of the tally correct.
As part of proving the boardroom voting contract correct we have mechanised the required results from their paper.

\sloppy An Ethereum version of the boardroom voting was developed by McCorry, Shahandashti and Hao~\cite{ethereum-boardroom}.
However, the contract is not formally verified.
Their version uses elliptic curves instead of finite fields to achieve the same security guarantees with much smaller key sizes and therefore more efficient computation.
Our contract uses finite fields and is less efficient.

Previous work in testing of smart contracts have been done in Echidna~\cite{grieco:Echidna}, Brownie, and ContractFuzzer~\cite{contractfuzzing}.
A common denominator for these works is that they choose a fuzzing approach where transactions are generated at random.
This leads to poor test coverage, and each work employs different automated methods to improve test coverage.
Unlike our testing framework, which allows for testing global properties about entire execution traces, these works only support testing assertional properties about single steps of execution. 

\section{Conclusion and Future Work}\label{sec:conclusions}
We have presented several extensions to the ConCert smart contract certification framework: certified extraction, integration of the ConCert execution model with QuickChick and two verified smart contracts (Escrow and Boardroom Voting) used as case studies for the developed techniques.
Currently, we support two target languages for smart contract extraction: Liquidity and Midlang.
Since Midlang is a derivative of the Elm programming language our extraction also allows targeting Elm.
Our extraction technique extends the certified erasure~\cite{CertErasure} and allows for targeting various functional smart contract languages.
Our experience shows that the extraction is well-suited for Coq programs in a fragment of Gallina that corresponds to a generic polymorphic functional language extended with refinement types.
This fragment is sufficient to cover most of the features found in functional smart contract languages.
In general, our pipeline allows for implementing, testing, verifying and extracting many interesting smart contracts, while retaining a small TCB.
Moreover, smart contracts like Escrow, Crowdfunding and the implementation of ERC-20 specification shows that the pipeline is suitable for real-world smart contracts.

We believe that with minor modifications of the Liquidity pretty-printer, we will be able to target the languages from the LIGO family by Tezos and other functional smart contract languages.
Moreover, targeting multi-paradigm languages with a functional subset is also possible.
We consider Rust as one of the future targets.
We plan to finalise the extraction of the boardroom voting contract so it performs well in the practical setting.
One way of achieving this would be to integrate it with extracted high-performance cryptographic primitives using the approach of FiatCrypto~\cite{FiatCrypto,Hvass:FieldInversion}.
We plan to extend the extraction of types to handle type schemes and improve handling inductives with no constructors (e.g.\ \icode{False}).
Our future work also includes adding more optimisation passes: removing singleton inductives (e.g.\ \icode{Acc}), expanding \icode{match} branches, and inlining definitions.
Some of these optimisations are necessary for extending the range of programs that can be extracted to target languages not supporting unsafe type coercions.

\section{Acknowledgments}
This work was partially supported by the Danish Industry Foundation in the Blockchain Academy Network project.

\newpage

\lstset{language=Caml}
\begin{appendices}
\crefalias{section}{appendix}
  \section{Extracted code for the \icode{counter} contract in Liquidity}\label{appendix:counter-liquidity}
  \begin{lstlisting}
let[@inline] fst (p : 'a * 'b) : 'a = p.(0)
let[@inline] snd (p : 'a * 'b) : 'b = p.(1)
let[@inline] addInt (i : int) (j : int) = i + j
let[@inline] subInt (i : int) (j : int) = i $-$ j
let[@inline] ltInt (i : int) (j : int) = i < j
type 'a sig_ = 'a
let exist_ a = a

type coq_msg = Coq_Inc of int | Coq_Dec of int
type coq_SimpleCallCtx = (timestamp * (address * (tez * tez)))
type storage = int
type coq_sumbool = Coq_left | Coq_right

let coq_my_bool_dec (b1 : bool) (b2 : bool) = (if b1 then fun x -> if x then Coq_left else Coq_right else fun x -> if x then Coq_right else Coq_left) b2

let coq_inc_counter (st : storage) (inc : ( (int) sig_)) =
  exist_ ((addInt st ((fun x -> x) inc)))

let coq_dec_counter (st : storage) (dec : ( (int) sig_)) =
  exist_ ((subInt st ((fun x -> x) dec)))

let coq_counter (msg : coq_msg) (st : storage) =
  match msg with
  | Coq_Inc i ->
   (match coq_my_bool_dec (ltInt 0 i) true with
    | Coq_left  -> Some ([],
      ((fun x -> x) (coq_inc_counter st (exist_ (i)))))
    | Coq_right  -> None)
  | Coq_Dec i ->
     (match coq_my_bool_dec (ltInt 0 i) true with
     | Coq_left  -> Some ([], ((fun x -> x) (coq_dec_counter st (exist_ (i)))))
     | Coq_right  -> None)

let%init storage (setup : int) =
  let inner (ctx : coq_SimpleCallCtx) (setup : int) = let ctx' = ctx in
  Some setup in
  let ctx = (Current.time (),
    (Current.sender (), (Current.amount (),Current.balance ()))) in
  match (inner ctx setup) with
  | Some v -> v
  | None -> failwith ()

let wrapper param (st : storage) =
  match coq_counter param st with
  | Some v -> v
  | None -> failwith ()

let%entry main param st = wrapper param st
  \end{lstlisting}

  \section{Extracted code for the \icode{counter} contract in Midlang}\label{appendix:counter-midlang}
  \begin{lstlisting}
import Basics exposing (..)
import Blockchain exposing (..)
import Bool exposing (..)
import Int exposing (..)
import Maybe exposing (..)
import Order exposing (..)
import Transaction exposing (..)
import Tuple exposing (..)

type Msg
  = Inc Int
  | Dec Int

type alias Storage = Int

type Option a
  = Some a
  | None

type Prod a b
  = Pair a b

type Sumbool
  = Left
  | Right

my_bool_dec : Bool -> Bool -> Sumbool
my_bool_dec b1 b2 =
  (case b1 of
     True ->
       \x -> case x of
               True ->
                 Left
               False ->
                 Right
     False ->
       \x -> case x of
               True ->
                 Right
               False ->
                 Left) b2

type Sig a
  = Exist a

proj1_sig : Sig a -> a
proj1_sig e =
  case e of
    Exist a ->
      a

inc_counter : Storage -> Sig Int -> Sig Storage
inc_counter st inc =
  Exist (add st (proj1_sig inc))

dec_counter : Storage -> Sig Int -> Sig Storage
dec_counter st dec =
  Exist (sub st (proj1_sig dec))

counter : Msg -> Storage -> Option (Prod Transaction Storage)
counter msg st =
  case msg of
    Inc i ->
      case my_bool_dec (lt 0 i) True of
        Left ->
          Some (Pair Transaction.none (proj1_sig (inc_counter st (Exist i))))
        Right ->
          None
    Dec i ->
      case my_bool_dec (lt 0 i) True of
        Left ->
          Some (Pair Transaction.none (proj1_sig (dec_counter st (Exist i))))
        Right ->
          None
  \end{lstlisting}
\end{appendices}

\bibliographystyle{alpha}
\bibliography{paper}

\newcommand{\etalchar}[1]{$^{#1}$}
\begin{thebibliography}{PCWD{\etalchar{+}}20}

\bibitem[AAM{\etalchar{+}}17]{Anand:CertiCoq}
Abhishek Anand, Andrew Appel, Greg Morrisett, Zoe Paraskevopoulou, Randy
  Pollack, Olivier Belanger, Matthieu Sozeau, and Matthew Weaver.
\newblock {CertiCoq: A verified compiler for Coq}.
\newblock In {\em CoqPL'2017}, {2017}.

\bibitem[AE18]{CertifiedCompCL}
Danil Annenkov and Martin Elsman.
\newblock Certified compilation of financial contracts.
\newblock In {\em PPDP'2018}, 2018.

\bibitem[AMNS20]{ConCertArtifact}
Danil Annenkov, Mikkel Milo, Jakob~Botsch Nielsen, and Bas Spitters.
\newblock {Source Code for Paper: Extracting Smart Contracts Tested and
  Verified in Coq}, 2020.

\bibitem[ANS20]{ConCert}
Danil Annenkov, Jakob~Botsch Nielsen, and Bas Spitters.
\newblock {ConCert: A Smart Contract Certification Framework in Coq}.
\newblock In {\em CPP'2020}, 2020.

\bibitem[BBE15]{CertFinContr}
Patrick Bahr, Jost Berthold, and Martin Elsman.
\newblock Certified symbolic management of financial multi-party contracts.
\newblock {\em SIGPLAN Not.}, 2015.

\bibitem[BG05]{CIC-type-decorations:Barras2005}
Bruno Barras and Benjamin Gr{\'e}goire.
\newblock On the role of type decorations in the calculus of inductive
  constructions.
\newblock In {\em CSL}, 2005.

\bibitem[BIL{\etalchar{+}}18]{Liquidity}
\c{C}agdas Bozman, Mohamed Iguernlala, Michael Laporte, Fabrice Le~Fessant, and
  Alain Mebsout.
\newblock {Liquidity: OCaml pour la Blockchain}.
\newblock In {\em {JFLA18}}, 2018.

\bibitem[BMM04]{Ind-falimilies-indices:Brady}
Edwin Brady, Conor McBride, and James McKinna.
\newblock Inductive families need not store their indices.
\newblock In Stefano Berardi, Mario Coppo, and Ferruccio Damiani, editors, {\em
  Types for Proofs and Programs}, pages 115--129. Springer Berlin Heidelberg,
  2004.

\bibitem[BN02]{Isabelle-ExecutingHOL}
Stefan Berghofer and Tobias Nipkow.
\newblock {Executing Higher Order Logic}.
\newblock In Paul Callaghan, Zhaohui Luo, James McKinna, Robert Pollack, and
  Robert Pollack, editors, {\em Types for Proofs and Programs}, pages 24--40,
  Berlin, Heidelberg, 2002. Springer Berlin Heidelberg.

\bibitem[BSH20]{Hvass:FieldInversion}
Bas~Spitters Benjamin S.~Hvass, Diego F.~Aranha.
\newblock {High-assurance field inversion forpairing-friendly primes}.
\newblock 2020.

\bibitem[CFL06]{ExecutingExtracted:CruzFilipeLetouzey}
Lu\'{\i}s Cruz-Filipe and Pierre Letouzey.
\newblock A large-scale experiment in executing extracted programs.
\newblock {\em Electron. Notes Theor. Comput. Sci.}, 2006.

\bibitem[CFS03]{ExtractionLargeProofs:CrusFilipeSpitters}
Lu{\'i}s Cruz-Filipe and Bas Spitters.
\newblock Program extraction from large proof developments.
\newblock In {\em Theorem Proving in Higher Order Logics}, 2003.

\bibitem[CH11]{claessen:QuickCheck}
Koen Claessen and John Hughes.
\newblock Quickcheck: a lightweight tool for random testing of haskell
  programs.
\newblock {\em Acm sigplan notices}, 46(4):53--64, 2011.

\bibitem[CKNW19]{Chapman:PlutusCore}
James Chapman, Roman Kireev, Chad Nester, and Philip Wadler.
\newblock {System F in Agda, for fun and profit}.
\newblock In {\em MPC'19}, 2019.

\bibitem[DEF18]{FairSwap}
Stefan Dziembowski, Lisa Eckey, and Sebastian Faust.
\newblock Fairswap: How to fairly exchange digital goods.
\newblock In {\em {ACM} Conference on Computer and Communications Security},
  pages 967--984. {ACM}, 2018.

\bibitem[EPG{\etalchar{+}}19]{FiatCrypto}
Andres Erbsen, Jade Philipoom, Jason Gross, Robert Sloan, and Adam Chlipala.
\newblock {Simple High-Level Code for Cryptographic Arithmetic - With Proofs,
  Without Compromises}.
\newblock In {\em {IEEE} Symposium on Security and Privacy}, 2019.

\bibitem[Fel20]{ElmInAction}
Richard Feldman.
\newblock {\em {Elm in Action}}.
\newblock 2020.

\bibitem[FL04]{FSets:FillaitreLetozey}
Jean-Christophe Filli{\^a}tre and Pierre Letouzey.
\newblock Functors for proofs and programs.
\newblock In David Schmidt, editor, {\em Programming Languages and Systems},
  pages 370--384. Springer Berlin Heidelberg, 2004.

\bibitem[GSC{\etalchar{+}}20]{grieco:Echidna}
Gustavo Grieco, Will Song, Artur Cygan, Josselin Feist, and Alex Groce.
\newblock Echidna: effective, usable, and fast fuzzing for smart contracts.
\newblock In {\em Proceedings of the 29th ACM SIGSOFT International Symposium
  on Software Testing and Analysis}, pages 557--560, 2020.

\bibitem[HLM20]{Henglein:CSL}
Fritz Henglein, Christian~Kj{\ae}r Larsen, and Agata Murawska.
\newblock A formally verified static analysis framework for compositional
  contracts.
\newblock In {\em Financial Cryptography and Data Security (FC)}, 2020.

\bibitem[HN07]{Isabelle-CodeGen}
Florian Haftmann and Tobias Nipkow.
\newblock {A code generator framework for Isabelle/HOL}.
\newblock In {\em Department of Computer Science, University of
  Kaiserslautern}, 2007.

\bibitem[HN18]{VerfiedCompIsabelle}
Lars Hupel and Tobias Nipkow.
\newblock A verified compiler from isabelle/hol to cakeml.
\newblock In Amal Ahmed, editor, {\em Programming Languages and Systems}, pages
  999--1026, 2018.

\bibitem[HRZ10]{open-vote-network}
Feng Hao, Peter~YA Ryan, and Piotr Zieli{\'n}ski.
\newblock Anonymous voting by two-round public discussion.
\newblock {\em IET Information Security}, 4(2), 2010.

\bibitem[JLC18]{contractfuzzing}
Bo~Jiang, Ye~Liu, and W.~K. Chan.
\newblock Contractfuzzer: Fuzzing smart contracts for vulnerability detection.
\newblock {\em CoRR}, abs/1807.03932, 2018.

\bibitem[KAE{\etalchar{+}}14]{klein14tocs}
Gerwin Klein, June Andronick, Kevin Elphinstone, Toby Murray, Thomas Sewell,
  Rafal Kolanski, and Gernot Heiser.
\newblock Comprehensive formal verification of an {OS} microkernel.
\newblock {\em ACM T. Comput. Syst.}, 32(1):2:1--2:70, 2014.

\bibitem[KMNO14]{CakeML}
Ramana Kumar, Magnus~O. Myreen, Michael Norrish, and Scott Owens.
\newblock {CakeML: A Verified Implementation of ML}.
\newblock In {\em Proceedings of the 41st ACM SIGPLAN-SIGACT Symposium on
  Principles of Programming Languages}, POPL '14, pages 179--191. ACM, 2014.

\bibitem[Kus17]{Kusee:agda-haskell}
W.~H. Kusee.
\newblock {Compiling Agda to Haskell with fewer coercions}, 2017.
\newblock Master's thesis.

\bibitem[Ler06]{2006-Leroy-compcert}
Xavier Leroy.
\newblock Formal certification of a compiler back-end, or: programming a
  compiler with a proof assistant.
\newblock In {\em POPL}, pages 42--54, 2006.

\bibitem[Let03]{CoqNewExtraction}
Pierre Letouzey.
\newblock A new extraction for coq.
\newblock In {\em Types for Proofs and Programs}, pages 200--219, 2003.

\bibitem[Let04]{letouzey04}
Pierre Letouzey.
\newblock {\em Programmation fonctionnelle certifi{\'e}e -- L'extraction de
  programmes dans l'assistant {Coq}}.
\newblock PhD thesis, Universit{\'e} Paris-Sud, 2004.

\bibitem[LST18]{Marlowe}
Pablo Lamela~Seijas and Simon Thompson.
\newblock Marlowe: Financial contracts on blockchain.
\newblock In Tiziana Margaria and Bernhard Steffen, editors, {\em International
  Symposium on Leveraging Applications of Formal Methods, Verification and
  Validation. Industrial Practice}, 2018.

\bibitem[LY98]{AlgM:LeeYi}
Oukseh Lee and Kwangkeun Yi.
\newblock Proofs about a folklore let-polymorphic type inference algorithm.
\newblock {\em ACM Trans. Program. Lang. Syst.}, 1998.

\bibitem[MPW{\etalchar{+}}18]{Oeuf}
Eric Mullen, Stuart Pernsteiner, James~R. Wilcox, Zachary Tatlock, and Dan
  Grossman.
\newblock {{\OE}uf: Minimizing the Coq Extraction TCB}.
\newblock In {\em CPP 2018}, 2018.

\bibitem[MSH17]{ethereum-boardroom}
Patrick McCorry, Siamak~F Shahandashti, and Feng Hao.
\newblock A smart contract for boardroom voting with maximum voter privacy.
\newblock In {\em FC 2017}, 2017.

\bibitem[NS19]{Interactions}
Jakob~Botsch Nielsen and Bas Spitters.
\newblock {Smart Contract Interactions in Coq}.
\newblock In {\em FMBC'2019}, 2019.

\bibitem[O'C17]{O'Connor:Simplicity}
Russell O'Connor.
\newblock {Simplicity: A New Language for Blockchains}.
\newblock PLAS17, 2017.

\bibitem[PCWD{\etalchar{+}}20]{ExtractionFiat}
Cl{\'e}ment Pit-Claudel, Peng Wang, Benjamin Delaware, Jason Gross, and Adam
  Chlipala.
\newblock Extensible extraction of efficient imperative programs with foreign
  functions, manually managed memory, and proofs.
\newblock In Nicolas Peltier and Viorica Sofronie-Stokkermans, editors, {\em
  Automated Reasoning}, pages 119--137, 2020.

\bibitem[PE16]{IdrisSmarContracts:Pettersson2016}
Jack Pettersson and Robert Edstr{\"o}m.
\newblock {Safer smart contracts through type-driven development}, 2016.
\newblock Master's thesis.

\bibitem[PHD{\etalchar{+}}15]{Paraskevopoulou:QuickChick}
Zoe Paraskevopoulou, Catalin Hritcu, Maxime D{\'{e}}n{\`{e}}s, Leonidas
  Lampropoulos, and Benjamin~C. Pierce.
\newblock Foundational property-based testing.
\newblock In Christian Urban and Xingyuan Zhang, editors, {\em 6th
  International Conference on Interactive Theorem Proving (ITP)}, volume 9236
  of {\em Lecture Notes in Computer Science}, pages 325--343. Springer, 2015.

\bibitem[SAB{\etalchar{+}}20]{MetaCoq}
Matthieu Sozeau, Abhishek Anand, Simon Boulier, Cyril Cohen, Yannick Forster,
  Fabian Kunze, Gregory Malecha, Nicolas Tabareau, and Th{\'e}o Winterhalter.
\newblock The metacoq project.
\newblock {\em Journal of Automated Reasoning}, Feb 2020.

\bibitem[SBF{\etalchar{+}}19]{CertErasure}
Matthieu Sozeau, Simon Boulier, Yannick Forster, Nicolas Tabareau, and Th\'{e}o
  Winterhalter.
\newblock {Coq Coq Correct! Verification of Type Checking and Erasure for Coq,
  in Coq}.
\newblock In {\em {POPL'2019}}, 2019.

\bibitem[SM19]{Equations}
Matthieu Sozeau and Cyprien Mangin.
\newblock Equations reloaded: High-level dependently-typed functional
  programming and proving in coq.
\newblock {\em Proc. ACM Program. Lang.}, 3(ICFP), 2019.

\bibitem[SNJ{\etalchar{+}}19]{Sergey:ScillaOOPSLA}
Ilya Sergey, Vaivaswatha Nagaraj, Jacob Johannsen, Amrit Kumar, Anton Trunov,
  and Ken Chan.
\newblock {Safer Smart Contract Programming with Scilla}.
\newblock In {\em OOPSLA19}, 2019.

\end{thebibliography}

\end{document}